\documentclass[11pt]{article}

\usepackage[preprint]{acl}

\usepackage{times}
\usepackage{latexsym}

\usepackage[T1]{fontenc}

\usepackage[utf8]{inputenc}

\usepackage{microtype}

\usepackage{inconsolata}

\usepackage{graphicx}
\usepackage{booktabs}
\usepackage{subcaption}
\usepackage{multirow}
\usepackage[table]{xcolor}
\usepackage[textsize=tiny]{todonotes}

\usepackage{svg}
\usepackage{amsmath}
\usepackage[capitalize,noabbrev]{cleveref}

\usepackage{listings}

\lstdefinestyle{prompt}{
  basicstyle=\scriptsize\ttfamily,
  breaklines=true,
  breakatwhitespace=true,
  columns=fullflexible,
  keepspaces=true,
  upquote=true,
  extendedchars=true,
  inputencoding=utf8,
  frame=single,
  framerule=0.5pt,
  rulecolor=\color{black!55},
  backgroundcolor=\color{gray!5},
  xleftmargin=2mm,
  xrightmargin=2mm,
  framexleftmargin=1mm,
  framexrightmargin=1mm,
  aboveskip=1ex,
  belowskip=1ex,
  literate=
    {—}{{---}}1
    {–}{{--}}1
    {‘}{{`}}1
    {’}{{'}}1
    {“}{{``}}1
    {”}{{''}}1
    {…}{{\ldots}}1
    {→}{{$\rightarrow$}}1
    {←}{{$\leftarrow$}}1
    {↑}{{$\uparrow$}}1
    {↓}{{$\downarrow$}}1
    {≤}{{$\leq$}}1
    {≥}{{$\geq$}}1
    {≠}{{$\neq$}}1
    {×}{{$\times$}}1
    {·}{{$\cdot$}}1
    {±}{{$\pm$}}1
    {°}{{$^\circ$}}1
    {α}{{$\alpha$}}1
    {β}{{$\beta$}}1
    {γ}{{$\gamma$}}1
    {δ}{{$\delta$}}1
    {ε}{{$\epsilon$}}1
    {λ}{{$\lambda$}}1
    {μ}{{$\mu$}}1
    {π}{{$\pi$}}1
    {σ}{{$\sigma$}}1
    {τ}{{$\tau$}}1
    {θ}{{$\theta$}}1
    {ω}{{$\omega$}}1
    {Δ}{{$\Delta$}}1
    {Σ}{{$\Sigma$}}1,
}

\title{\textsc{VEX-Bench}: 
Benchmarking LLM Agents for Assessing Exploitability of Software Supply Chain Vulnerabilities
}

\author{
 \textbf{Jiahao Shi\textsuperscript{1}},
 \textbf{Edward Tsien\textsuperscript{2}},
 \textbf{Yifeng Di\textsuperscript{1}},
 \textbf{Hongjiao Zhang\textsuperscript{2}},
\\
\textbf{Yuan Tang\textsuperscript{2}},
 \textbf{Ronit Dey\textsuperscript{2}},
 \textbf{Ilona Shishov\textsuperscript{2}},
 \textbf{Gal Netanel \textsuperscript{2}},
 \textbf{Zvi Grinberg\textsuperscript{2}},
\\
 \textbf{Vladimir Belousov\textsuperscript{2}},
 \textbf{Bat-Zion Rotman\textsuperscript{2}},
 \textbf{Ilan Pinto\textsuperscript{2}},
 \textbf{Tianyi Zhang\textsuperscript{1}}
\\
 \textsuperscript{1}Department of Computer Science, Purdue University \\
 \textsuperscript{2}Red Hat
\\
 \texttt{\{shi768, tianyi\}@purdue.edu} \\
 \texttt{\{etsien\}@redhat.com}
}

\begin{document}
\maketitle

\begin{abstract}
The software supply chain has become an increasingly exposed attack surface because of its reliance on intricate yet fragile dependencies. Existing defenses such as GitHub Dependabot often raise many false alerts because their coarse-grained matching cannot determine whether a vulnerable dependency is actually exploitable. Security analysts typically spend substantial time assessing vulnerability exploitability case by case. Recent LLM agents have emerged as promising candidates for this task given their advanced capabilities in coding and cybersecurity, yet no existing benchmark evaluates them on it. Prior benchmarks target zero-day settings, where agents detect and exploit previously unknown vulnerabilities. In contrast, software supply chain security focuses on how known vulnerabilities in upstream dependencies affect downstream projects. This requires agents to reason across repositories and determine whether an upstream vulnerability is exploitable in the downstream project.
To address this gap, we introduce \textsc{VEX-Bench}, the first benchmark for evaluating LLM agents' ability to assess the exploitability of software supply chain vulnerabilities. It contains 75 real-world cases mined from GitHub and labeled by security experts, covering Python, Java, and Go. We evaluate nine models across three agent harnesses. While GPT-5.5 and Claude Opus 4.6 reach approximately 80\% F1 on binary vulnerability-status classification, only GPT-5.5 surpasses 70\% macro-F1 on fine-grained justification classification. This gap highlights the challenge of moving beyond binary exploitability assessment to identifying fine-grained exploitability reasons. Code and data:~\url{https://github.com/steven1518/vex-bench}
\end{abstract}

\section{Introduction}
The software supply chain faces increasingly severe security threats. Sonatype reports that software supply chain attacks have grown 156\% year-over-year~\citep{sonatype2024sscr}, and the 2025 OWASP Top 10 ranks \emph{Software Supply Chain Failures} third overall, with 50\% of surveyed practitioners ranking it their top concern~\citep{owasp2025supplychain}. A single vulnerability, Log4Shell~\citep{log4shell2021}, affected hundreds of millions of devices and 93\% of enterprise cloud environments, causing billions of dollars in industry-wide remediation costs.
This growing severity stems largely from modern software's reliance on expansive package ecosystems (e.g., npm, PyPI, Maven) to reduce development costs and accelerate delivery. This creates complex dependencies among packages, allowing a single vulnerable package to propagate risk into a vast number of downstream projects.

To defend against such risks, existing tools such as GitHub Dependabot~\citep{dependabot} scan software components against vulnerability databases to identify known vulnerabilities. However, such coarse-grained matching produces a large number of false alerts, as it relies primarily on package names and versions, without considering contextual information or fine-grained reachability analysis~\citep{DBLP:conf/ndss/PuYG26, 9201023}. Security analysts are often overwhelmed by these alerts and spend substantial effort triaging them, leading to alert fatigue and causing critical vulnerabilities to be buried among spurious ones~\citep{10.1109/TSE.2023.3278129, 10.1145/3522587, zhou2025reality}.
These limitations motivate leveraging LLM agents, whose strong coding and cybersecurity capabilities~\citep{anthropic2026opus46, openai2026gpt55} make them well-suited for assessing exploitability of software supply chain vulnerabilities.

However, existing cybersecurity benchmarks for evaluating LLM agents~\citep{zhang2026bountybench, wang2026cybergym, lee2026secbench, zhu2025cvebench, zhang2025cybench} focus on zero-day settings, where agents discover or exploit previously unknown vulnerabilities within a single codebase. In contrast, software supply chain security poses a completely different setting---the vulnerability is already known and is introduced into the local project through a third-party dependency. This task therefore requires agents to integrate external vulnerability evidence with cross-repository reasoning to assess downstream exposure.

To address this gap, we construct \textsc{\textbf{VEX-Bench}}, the first benchmark for evaluating LLM agents' ability to assess the exploitability of software supply chain vulnerabilities. \textsc{VEX-Bench} contains 75 real-world cases mined from GitHub across Go, Python, and Java projects, with ground-truth labels manually annotated by security experts. Given a codebase and a known vulnerability in one of its dependencies, an agent must determine whether the codebase is genuinely affected. Beyond the binary affected/not-affected decision, we define fine-grained justification labels that explain why a downstream project is not affected, mirroring how vulnerability status is communicated in industry. These choices together enable a more realistic evaluation of LLM agents on real-world software supply chain tasks.

We evaluate nine models across three agent harnesses on \textsc{VEX-Bench}. For binary vulnerability-status classification, Claude Opus~4.6 and GPT-5.5 achieve the highest status F1 scores, reaching 81.6\% and 79.9\%, respectively. Performance drops at the finer justification granularity: only GPT-5.5 exceeds 70\% macro-F1. This indicates that identifying the precise exploitability reason is more challenging than making the coarse affected/not-affected decision. We further analyze performance across inference cost, programming language, repository size, and harness choice, providing a more detailed picture of where fine-grained exploitability reasoning remains difficult.

\section{Related Work}

\subsection{Software Supply Chain Security}
\label{sec:soft-sup-chain-sec}

Many known vulnerabilities propagate risk to downstream projects through software supply chains~\citep{10172868, shen2025understanding}. Software composition analysis (SCA) tools automatically identify third-party components in a project and flag vulnerable dependencies by matching package names and versions against vulnerability databases~\citep{ponta2020detection, imtiaz2021comparative, zhao2023software}. However, such metadata-based matching is often imprecise and generates many false alerts~\citep{9201023, DBLP:conf/ndss/PuYG26}.
Another line of work uses reachability analysis to filter alerts whose vulnerable functions are never invoked~\citep{ponta2020detection, jia2025cpvra}. However, reachability analysis still faces several limitations. Identifying the exact vulnerable functions is itself challenging~\citep{dunlap2024pairing}. Call graphs are difficult to make both sound and precise. Most importantly, reachability does not imply exploitability, which may additionally depend on configuration and environmental conditions.

In this work, we propose \textsc{VEX-Bench} to evaluate whether LLM agents can perform end-to-end exploitability assessments of vulnerable dependencies in downstream projects.

\subsection{Cybersecurity Benchmarks for Agents}
Recent benchmarks evaluate LLM agents on cybersecurity tasks from two main angles. CyBench~\citep{zhang2025cybench} uses Capture the Flag (CTF) challenges, where agents operate in compact, self-contained environments and recover hidden flags by exploiting intentionally planted vulnerabilities. Other benchmarks use real-world software projects, drawing from historical CVEs, OSS-Fuzz reports, or bug-bounty disclosures to ask agents to detect, exploit, or patch vulnerabilities in the repositories that contain the flawed code~\citep{wang2026cybergym, lee2026secbench, zhu2025cvebench, zhang2026bountybench}.

These benchmarks primarily focus on \emph{first-party} vulnerabilities within a single codebase, overlooking software supply chain vulnerabilities. \textsc{VEX-Bench} instead studies dependency-level exploitability assessment: given a downstream project and a known CVE in one of its third-party dependencies, the agent must determine whether the project is actually affected by searching for upstream vulnerability information and inspecting downstream dependency usage.

\begin{figure*}[t]
    \centering
    \includegraphics[width=0.9\linewidth]{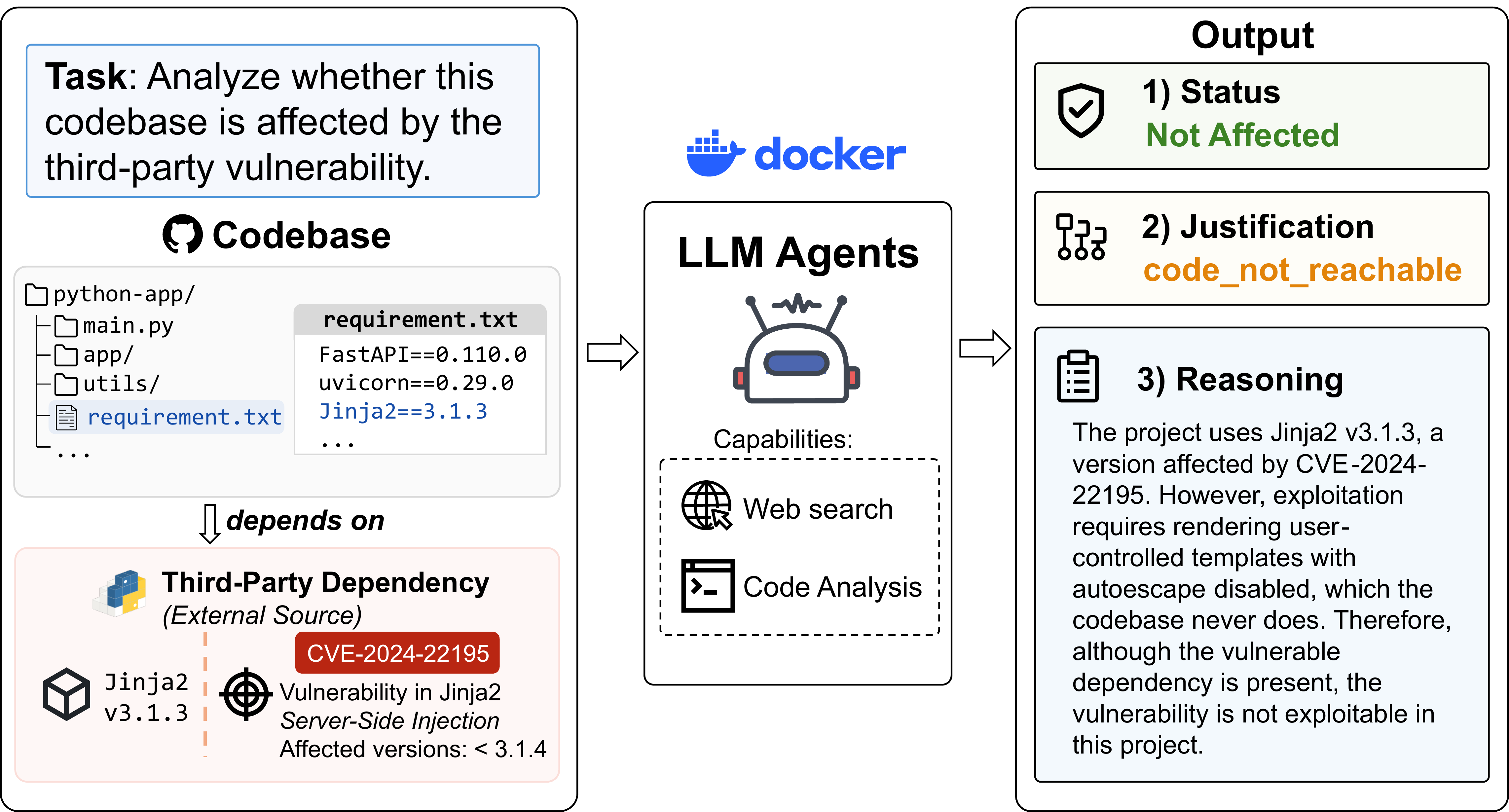}
    \caption{\textsc{VEX-Bench} sources task instances from real-world projects by pairing codebases with third-party dependency vulnerabilities (CVEs) to determine real-world exploitability. Provided with the project's source code and CVE identifier, LLM agents search external information and analyze code to produce a vulnerability status (Affected / Not Affected), a justification label (e.g., code\_not\_reachable), and reasoning grounded in the codebase.}
    \label{fig:vex-bench}
\end{figure*}

\section{\textsc{VEX-Bench}}

We construct \textsc{VEX-Bench} to evaluate whether LLM-based agents can determine if a project is genuinely affected by a known vulnerability in its dependencies. The task requires agents to analyze the local codebase, retrieve information from external sources, inspect upstream projects and vulnerability reports, and synthesize these signals into an end-to-end exploitability analysis. This assessment is routinely performed by security analysts, and its results are communicated through the Vulnerability Exploitability eXchange (VEX) format~\citep{ntia2021vex}, a machine-readable specification for reporting whether a known vulnerability actually affects a given product.

\subsection{Task Formulation}
\label{sec:task-formulation}
We formalize the task as cross-repository vulnerability exploitability assessment. As shown in \Cref{fig:vex-bench}, the agent operates within a target codebase as its working environment and is prompted with a CVE identifier for a known vulnerability in dependencies. The agent must analyze the codebase against the vulnerability and produce a structured prediction together with supporting evidence.

\paragraph{Input Formulation}
Each task consists of a \emph{target codebase} and a \emph{vulnerability identifier}. The codebase is a complete real-world project that serves as the agent's working environment, in which it reads files, runs commands, and analyzes code. The vulnerability identifier is a CVE referring to a known issue in one of the codebase's third-party dependencies, and is delivered to the agent as part of its instruction prompt. The full prompt template is provided in Appendix~\ref{app:task-inst-prompt}.

By design, the agent receives no curated advisory text, no preselected vulnerable files, and no pre-computed call graph slice. It must retrieve information about the vulnerability from external sources and locate the relevant code paths in the codebase without assistance. This mirrors the situation a human expert faces when a scanner flags a production repository, and ensures that our evaluation reflects the agent's own cross-source reasoning rather than the leverage of human-curated hints.

\paragraph{Output Formulation}

The agent produces a structured output with three fields. The \emph{vulnerability status} is a binary label indicating whether the codebase is affected. When the status is not-affected, the \emph{justification} further specifies the reason, drawn from four fine-grained not-affected categories. The \emph{reasoning trace} is a free-form explanation recording which call sites the agent inspected, which input paths it followed, and which mitigations it identified — making the prediction auditable rather than a black-box verdict. Together, these fields support both quick triage on the binary status and deeper review through the justification and trace.

\paragraph{Justification Labels}
Although the exploitability assessment is binary at the top level, a single \texttt{not\_affected} status can arise from different reasons. In one case the vulnerable function may never be invoked along any reachable call path; in another, the exploit may require preconditions that the project's default configuration prevents. Distinguishing these cases serves two purposes. First, it has direct practical value, since different reasons imply different follow-up actions for the downstream maintainer, and the specific reason an agent identifies reveals which kind of reasoning it is actually performing. Second, it is necessary for reliable evaluation. \texttt{not\_affected} cases dominate real-world triage outcomes, so the dataset is heavily imbalanced and a binary status prediction is easy to inflate by simply predicting the majority class.
Our schema is adapted from the VEX standard~\citep{cisa2022vexstatus}. \textsc{VEX-Bench} defines four justification categories that capture the most common reasons a known vulnerability fails to be exploitable in a downstream project. Detailed definitions and examples for each category appear in Appendix~\ref{appendix:justification}.

\subsection{Benchmark Construction}
We construct \textsc{VEX-Bench} through three stages: (1) mining candidate cases from GitHub, (2) rule-based filtering, and (3) manual annotation.

\paragraph{Mining GitHub Data}
We begin by targeting popular and mature open-source projects on GitHub, as such projects better reflect real-world software complexity, active maintenance, and practical dependency management scenarios. Specifically, we select the top 100 starred repositories with permissive open-source licenses (MIT, Apache 2.0, BSD 2-Clause, and BSD 3-Clause). From these repositories, we collect all pull requests (PRs) created after 2021 whose titles or descriptions reference a CVE identifier, yielding a coarse-grained pool of potential dependency-level vulnerabilities.

\paragraph{Rule-based Filtering}
Because the initial candidate pool contains irrelevant cases, such as release notes and internal project vulnerabilities, we apply heuristic filters to remove low-quality candidates.
Specifically, we first exclude cases in which all modified files are under test or example directories. Second, we exclude PRs that show more than five distinct CVEs. Finally, the most decisive rule requires the PR to modify a dependency manifest or lockfile (e.g., \texttt{go.mod}, \texttt{requirements.txt}, \texttt{setup.py}, \texttt{pom.xml}), which provides direct evidence that the host project actually consumes and reacts to the referenced vulnerable dependency. The PRs surviving this stage form the candidate pool for manual annotation.

\paragraph{Manual Annotation}

The filtered candidates still require manual labeling, as no reliable automatic signal can serve as ground truth. Existing automated approaches are either inaccurate or require substantial manual effort, as discussed in \Cref{sec:soft-sup-chain-sec}. Moreover, PR statuses on GitHub (e.g., merged or closed) cannot be directly interpreted as exploitability labels, since such decisions are often influenced by compatibility concerns or maintenance trade-offs. We therefore opt for manual annotation throughout.

Five annotators with prior experience in software engineering and security manually analyze and label the candidate PRs, with each case taking roughly 30 minutes to annotate. To ensure inter-annotator consistency, we adopt a calibration-then-scaling protocol. In the calibration phase, all five annotators independently label the same 15 cases drawn across the three languages, achieving an initial Fleiss' Kappa of 0.667. They then meet to reconcile disagreements and align on labeling criteria. In the scaling phase, annotators label distinct sets of cases under the calibrated guidelines, and each annotation is reviewed by at least one additional annotator. During review, 88.3\% of the initial labels are confirmed without revision, while the remaining disagreements are resolved through discussion until consensus is reached. This process yields a final benchmark of 75 cases spanning Go, Python, and Java.

\subsection{Benchmark Statistics}
\label{sec:benchmark-statistics}

\begin{table}[th]
\centering
\caption{This table reports the composition and codebase size of \textsc{VEX-Bench}, covering 75 cases across 67 CVEs and 35 real-world projects in Go, Python, and Java, where LOC denotes lines of code and Files denotes the number of source files in the target codebase.}
\label{tab:benchmark-stats}
\resizebox{\columnwidth}{!}{%
  \begin{tabular}{lrrrrr}
  \toprule
  \textbf{Language} & \textbf{Cases} & \textbf{CVEs} & \textbf{Projects} & \textbf{Median LOC} & \textbf{Median Files} \\
  \midrule
  Java   & 25 & 23 & 10 & 545{,}806 & 6{,}494 \\
  Go     & 30 & 24 & 15 &  90{,}013 &   924 \\
  Python & 20 & 20 & 10 & 165{,}122 & 2{,}124 \\
  \midrule
  All    & 75 & 67 & 35 & 272{,}791 & 2{,}671 \\
  \bottomrule
  \end{tabular}}\end{table}

\begin{figure}[th]
\centering
\includegraphics[width=\linewidth]{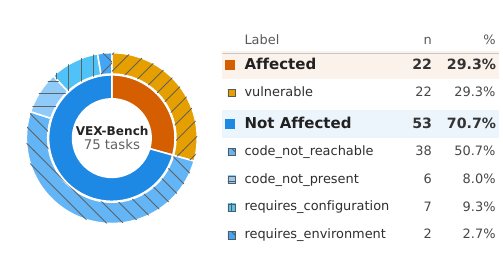}
\caption{Label distribution of \textsc{VEX-Bench} across 75 cases, where the inner ring shows the binary vulnerability status (affected vs.\ not affected) and the outer ring breaks down each status into its fine-grained justification category.}
\label{fig:label-dist}
\end{figure}

\textsc{VEX-Bench} contains 75 cases covering 67 CVEs drawn from 35 real-world repositories across Go, Python, and Java, as shown in \Cref{tab:benchmark-stats}. The target codebases are full projects, not curated snippets. The median case spans roughly 273K lines of code across more than 2,600 files, and the Java subset is the largest, with a median of 546K LOC. An agent therefore cannot solve the task by scanning the entire project; it must locate the few dependency call sites relevant to the vulnerability within a large codebase. The cases also vary widely in scale, ranging from 9,902 LOC to about 4.0M, which prevents strong performance from being explained by overfitting to a single repository size.

Most scanner-reported dependency vulnerabilities in \textsc{VEX-Bench} are not exploitable in the downstream project. \Cref{fig:label-dist} shows that only 29.3\% of cases are genuinely \texttt{affected}, while 70.7\% are not. Within the \texttt{not\_affected} group, \texttt{code\_not\_reachable} alone accounts for 38 cases, or 50.7\% of the entire benchmark and 71.7\% of all \texttt{not\_affected} cases. The remaining \texttt{not\_affected} cases are split across configuration, missing-code, and environment reasons. This long-tailed structure motivates fine-grained labels: a single \texttt{not\_affected} status would conflate reachability analysis with configuration, environment, and dependency reasoning, all of which require distinct agent capabilities.

\subsection{Defense Against Benchmark Hacking}

As LLM agents become more capable and are granted broader execution privileges, recent work has shown that they can exploit weaknesses in benchmark environments, such as accessing hidden labels or tampering with evaluation logic to inflate reported scores~\citep{thaman2026reward, atinafu2026rewardhackingagents}. To mitigate these risks, we execute all experiments inside isolated Docker environments. Each container is separated from the benchmark labels and evaluation artifacts, ensuring that agents can only solve tasks through reasoning rather than by accessing ground-truth answers. In addition, our evaluation does not rely on executable test cases, eliminating the possibility of test-harness manipulation. To further prevent leakage through version history, we remove Git history and provide agents only with a standalone snapshot of the target commit. The result is a sealed environment with no path to the answer, and it is identical across runs, preserving reproducibility.

Another potential source of leakage is web access. Since agents can search the web, they could in principle attempt to retrieve task-specific information online. However, the benchmark ground truth is not publicly available: all vulnerability-status labels and justifications are manually annotated by our security experts rather than derived from existing online sources. An agent may therefore gather background information about a vulnerability, but it cannot directly recover the benchmark answers. We further examine potential model-side leakage from the source PRs and find no meaningful evidence that they expose the benchmark answers or that the evaluated models memorized their contents. The full analysis is provided in Appendix~\ref{app:model-side-leakage}.

\section{Evaluation}

\begin{table*}[!t]
\centering
\caption{Performance of agents on \textsc{VEX-Bench}, reported as mean$\pm$standard deviation over three runs. Vulnerability~status metrics treat the task as binary classification; justification metrics evaluate fine-grained reasoning. Token and cost columns report per-case averages. Best and second-best mean values are \textbf{bolded} and \underline{underlined}, respectively; lower is better for Tokens/Cost.}
\label{tab:main-results}
\small
\setlength{\tabcolsep}{4pt}
\renewcommand{\arraystretch}{1.1}
\definecolor{harnessRow}{RGB}{225, 232, 240}
\definecolor{statusCol}{RGB}{240, 247, 240}
\definecolor{justifyCol}{RGB}{249, 242, 230}
\newcommand{\meanstd}[2]{#1{\scriptsize\textcolor{black!60}{$\pm$#2}}}
\resizebox{\textwidth}{!}{%
\begin{tabular}{l>{\columncolor{statusCol}}l>{\columncolor{statusCol}}l>{\columncolor{statusCol}}l>{\columncolor{statusCol}}l>{\columncolor{justifyCol}}l>{\columncolor{justifyCol}}l>{\columncolor{justifyCol}}lll}
\toprule
\multirow{2}{*}{\textbf{Model}} & \multicolumn{4}{c}{\textbf{Status}} & \multicolumn{3}{c}{\textbf{Justification}} & \multirow{2}{*}{\textbf{Tokens}} & \multirow{2}{*}{\textbf{Cost}} \\
\cmidrule(lr){2-5} \cmidrule(lr){6-8}
 & \multicolumn{1}{>{\columncolor{white}}c}{Accuracy}
 & \multicolumn{1}{>{\columncolor{white}}c}{Precision}
 & \multicolumn{1}{>{\columncolor{white}}c}{Recall}
 & \multicolumn{1}{>{\columncolor{white}}c}{F1}
 & \multicolumn{1}{>{\columncolor{white}}c}{Accuracy}
 & \multicolumn{1}{>{\columncolor{white}}c}{Macro-F1}
 & \multicolumn{1}{>{\columncolor{white}}c}{Weighted-F1}
 & & \\
\midrule
\rowcolor{harnessRow}
\multicolumn{10}{l}{\textit{w/ Claude Code}} \\
Claude Opus 4.6   & \meanstd{\underline{88.4}}{2.0} & \meanstd{76.3}{1.4} & \meanstd{\textbf{87.9}}{6.9} & \meanstd{\textbf{81.6}}{3.8} & \meanstd{\underline{76.4}}{2.8} & \meanstd{61.1}{3.4} & \meanstd{\underline{78.2}}{3.0} & 297.47K & \$0.595 \\
Claude Sonnet 4.6 & \meanstd{81.8}{5.4} & \meanstd{68.4}{7.1} & \meanstd{\underline{75.8}}{10.5} & \meanstd{71.8}{8.5} & \meanstd{67.1}{5.4} & \meanstd{56.1}{2.6} & \meanstd{70.5}{4.8} & 276.24K & \$0.281 \\
\midrule
\rowcolor{harnessRow}
\multicolumn{10}{l}{\textit{w/ Codex}} \\
GPT-5.5           & \meanstd{\textbf{89.3}}{4.0} & \meanstd{\textbf{88.7}}{6.2} & \meanstd{72.7}{9.1} & \meanstd{\underline{79.9}}{8.0} & \meanstd{\textbf{78.2}}{3.9} & \meanstd{\textbf{73.5}}{3.3} & \meanstd{\textbf{81.5}}{3.3} & 700.28K & \$1.082 \\
GPT-5.4 mini      & \meanstd{82.2}{3.9} & \meanstd{\underline{79.1}}{3.7} & \meanstd{53.0}{13.1} & \meanstd{63.1}{10.3} & \meanstd{66.7}{4.0} & \meanstd{60.4}{3.3} & \meanstd{71.9}{3.7} & 947.25K & \$0.274 \\
\midrule
\rowcolor{harnessRow}
\multicolumn{10}{l}{\textit{w/ OpenCode}} \\
Kimi K2.6         & \meanstd{81.8}{2.0} & \meanstd{74.5}{2.0} & \meanstd{66.7}{6.9} & \meanstd{70.3}{4.8} & \meanstd{72.9}{4.3} & \meanstd{\underline{64.6}}{7.9} & \meanstd{75.8}{4.1} & 579.01K & \$0.222 \\
DeepSeek-V4-Pro   & \meanstd{84.0}{6.1} & \meanstd{74.6}{9.9} & \meanstd{68.2}{13.6} & \meanstd{71.2}{11.9} & \meanstd{76.0}{7.4} & \meanstd{60.3}{5.1} & \meanstd{77.3}{6.2} & \textbf{170.60K} & \$0.075 \\
DeepSeek-V4-Flash & \meanstd{72.9}{5.1} & \meanstd{62.7}{11.3} & \meanstd{40.9}{0.0} & \meanstd{49.3}{3.4} & \meanstd{58.7}{5.8} & \meanstd{47.2}{3.3} & \meanstd{62.8}{3.8} & \underline{189.72K} & \textbf{\$0.027} \\
MiniMax-M2.7      & \meanstd{71.1}{10.1} & \meanstd{75.1}{1.8} & \meanstd{54.6}{12.0} & \meanstd{62.7}{7.7} & \meanstd{59.1}{13.5} & \meanstd{44.8}{12.0} & \meanstd{64.5}{11.1} & 245.28K & \underline{\$0.037} \\
GLM-5.1           & \meanstd{84.4}{4.1} & \meanstd{77.8}{6.8} & \meanstd{68.2}{7.9} & \meanstd{72.5}{6.2} & \meanstd{74.7}{4.8} & \meanstd{61.7}{3.2} & \meanstd{77.8}{3.6} & 201.45K & \$0.095 \\
\bottomrule
\end{tabular}
}
\end{table*}

\subsection{Experimental Setup}
\paragraph{Models and Harness}

Modern agents usually have two parts: a \emph{model} that performs reasoning and a \emph{harness} that manages execution, memory, tool use, and interaction with the environment. We therefore evaluate agents on \textsc{VEX-Bench} from both perspectives.

We select nine models, covering both closed-source and open-weight models. The closed-source group includes Claude Opus~4.6 \citep{anthropic2026opus46}, Claude Sonnet~4.6 \citep{anthropic2026sonnet46}, GPT-5.5 \citep{openai2026gpt55}, and GPT-5.4~mini \citep{openai2026gpt54mini}. The open-weight group includes five recent models reported to be competitive on agentic and coding benchmarks: Kimi~K2.6 \citep{moonshot2026kimik26}, DeepSeek-V4-Pro and DeepSeek-V4-Flash \citep{deepseek2026v4}, MiniMax-M2.7 \citep{minimax2026m27}, and GLM-5.1 \citep{zai2026glm51}.

For harnesses, we evaluate both vendor-specific and open-source frameworks. Vendor-specific harnesses are paired with their corresponding models: Claude Code~\citep{anthropic2025claudecode} with Claude Opus~4.6 and Claude Sonnet~4.6, and Codex~\citep{openai2025codex} with GPT-5.5 and GPT-5.4~mini. We evaluate the open-weight models with OpenCode~\citep{sst2025opencode}, a widely used open-source agent harness. Details of these models and frameworks are in Appendix~\ref{app:experimental-details}. Unless otherwise stated, model names below refer to the model under its corresponding default harness.

\paragraph{Metrics}
We use classification metrics to assess predictions of vulnerability status and justification. The vulnerability status is a binary label indicating whether the codebase is affected by the vulnerability. We report standard binary classification metrics, including accuracy, precision, recall, and F1 score. The justification label is a multi-class label over the affected class and the four not-affected reasons, evaluating the agent's fine-grained reasoning beyond the binary status prediction. We report accuracy, macro-F1, and weighted-F1 for this task.

\paragraph{Other Settings}
To ensure reproducibility and prevent benchmark leakage, all experiments are conducted inside isolated Docker environments. We build separate Linux-based Docker images for different programming languages, each equipped with the corresponding build toolchains. The environment details are in Appendix~\ref{app:experimental-details}.

The task prompt contains a broader set of categories than the five annotated
benchmark labels. Predictions outside the annotated label space are treated as
\texttt{FAILED}. The complete prompt and label-mapping rules are provided in
Appendix~\ref{app:task-inst-prompt}.

We evaluate each model--harness configuration three times and report the mean and standard deviation across runs. To prevent excessively long or non-terminating executions, we impose a maximum runtime limit of 10 minutes per task. Executions that exceed this limit are terminated and recorded as timeouts. Overall, 98.2\% of executions finish within this limit; timeouts are counted as failures and concentrate in MiniMax-M2.7 (12.0\%) and Kimi-K2.6 (3.1\%). Detailed runtime statistics are provided in Appendix~\ref{app:runtime-analysis}.

\subsection{Main Results}

\Cref{tab:main-results} reports the performance of nine model and harness configurations. While current agents perform well on the coarse vulnerability-status task, they remain far from saturated on fine-grained justification. GPT-5.5 obtains the best vulnerability-status accuracy at 89.3\%, followed by Claude Opus~4.6 at 88.4\%. Claude Opus~4.6 achieves the best status recall and F1 score, reaching 87.9\% recall and 81.6\% F1, while GPT-5.5 obtains the best status precision at 88.7\%. GPT-5.5 also leads all three justification metrics, reaching 78.2\% accuracy, 73.5\% macro-F1, and 81.5\% weighted-F1. At the lower end, MiniMax-M2.7 has the lowest status accuracy at 71.1\% and the lowest justification macro-F1 at 44.8\%. DeepSeek-V4-Flash records the lowest status precision, recall, and F1 score, as well as the lowest justification accuracy and weighted-F1, reaching 58.7\% justification accuracy and 62.8\% weighted-F1.

\textbf{Performance consistently drops from vulnerability-status prediction to justification prediction.} For all nine configurations, justification accuracy is lower than status accuracy. The largest drop occurs for GPT-5.4~mini, from 82.2\% status accuracy to 66.7\% justification accuracy, a gap of 15.5 percentage points. GPT-5.5 and Claude Opus~4.6 also show clear drops of 11.1 and 12.0 points, respectively. Thus, reporting only the binary affected/not-affected status can overstate performance relative to the finer-grained reasoning needed for software supply-chain triage.

\subsection{Comparison with Traditional SCA Tools}

\begin{table}[t]
\centering
\small
\caption{Performance of traditional SCA baselines on \textsc{VEX-Bench}. Cov., Acc., Prec., and Rec. denote coverage, accuracy, precision, and recall, respectively. Coverage is the percentage of repositories successfully analyzed. OSV (detect.) uses OSV-Scanner for dependency detection, whereas OSV (reach.) enables its vulnerable-function reachability analysis. The full benchmark contains all 75 cases, and the Go subset contains 30 cases. All values are percentages.}
\label{tab:sca-baselines}
\begin{tabular*}{\columnwidth}{@{\extracolsep{\fill}}lccccc@{}}
\toprule
\textbf{Method} & \textbf{Cov.} & \textbf{Acc.} & \textbf{Prec.} & \textbf{Rec.} & \textbf{F1} \\
\midrule
\multicolumn{6}{l}{\textit{Full benchmark ($n=75$)}} \\
OSV-Scanner & 93.3 & 33.3 & 27.4 & 77.3 & 40.5 \\
Trivy & 96.0 & 38.7 & 26.0 & 59.1 & 36.1 \\
\midrule
\multicolumn{6}{l}{\textit{Go subset ($n=30$)}} \\
OSV (detect.) & 100.0 & 26.7 & 24.1 & 100.0 & 38.9 \\
OSV (reach.) & 100.0 & 46.7 & 28.6 & 85.7 & 42.9 \\
govulncheck & 63.3 & 36.7 & 22.2 & 28.6 & 25.0 \\
\bottomrule
\end{tabular*}
\end{table}

To assess the performance of traditional software composition analysis (SCA) tools, we select Trivy~\cite{trivy} and OSV-Scanner~\cite{osvscanner}, two widely used tools, as baselines on the full dataset. We further evaluate OSV-Scanner's reachability mode and govulncheck~\cite{govulncheck} on the 30 Go cases, as reachability analysis requires mappings from CVEs to vulnerable functions, and reliable function-level records are available only for Go. We predict \texttt{affected} if a tool reports the target CVE and \texttt{not affected} otherwise.

\Cref{tab:sca-baselines} reports the performance of traditional SCA baselines on \textsc{VEX-Bench}. Traditional SCA tools can flag a broad set of potential vulnerabilities, but their low precision shows that they cannot reliably distinguish affected projects from false alerts. Reachability analysis does not close this gap. Govulncheck requires the target project to build successfully and therefore completes only 19 of the 30 cases. 
Moreover, reachability tools determine whether a vulnerable function appears in the call graph, but this does not establish whether the corresponding path is feasible or whether its branch conditions can actually trigger the vulnerable behavior. This limitation, together with the difficulty of constructing call graphs that are both sound and precise, explains why 70\% of their false positives are still labeled \texttt{code\_not\_reachable} and prevents reachability analysis from reliably assessing exploitability.

\subsection{Analysis Results}

\subsubsection{Cost-Performance Analysis}

\begin{figure}[!t]
    \centering
    \includegraphics[width=\linewidth]{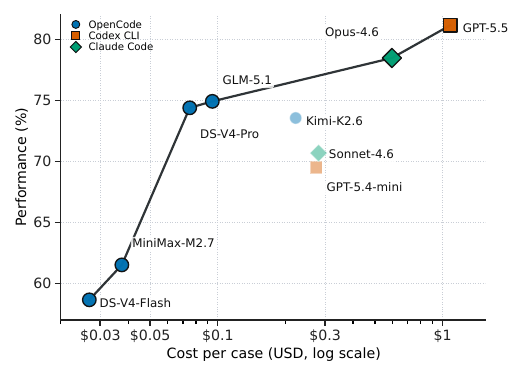}
    \caption{Performance versus per-case inference cost on \textsc{VEX-Bench}. The x-axis uses a log scale, and the line connects Pareto-efficient configurations.}
    \label{fig:perf-vs-cost}
\end{figure}
\textbf{Performance generally improves with inference cost, but the gains show diminishing returns.} \Cref{fig:perf-vs-cost} plots per-case cost against a balanced performance score \(P=\frac{1}{2}((A_s+F_s)/2+(A_j+M_j+W_j)/3)\), where \(A_s\) and \(F_s\) denote status accuracy and status F1, and \(A_j\), \(M_j\), and \(W_j\) denote justification accuracy, macro-F1, and weighted-F1. GPT-5.5 achieves the highest performance at 81.2\%, but is also the most expensive configuration at \$1.082 per case. Claude Opus~4.6 follows at 78.5\% and costs \$0.595 per case. In contrast, DeepSeek-V4-Pro reaches 74.4\% at only \$0.075 per case, and GLM-5.1 reaches 74.9\% at \$0.095 per case. These results suggest that inference cost is an incomplete proxy for exploitability-analysis quality: GPT-5.5 and Claude Opus~4.6 are strongest in absolute performance, while DeepSeek-V4-Pro and GLM-5.1 provide better cost-performance trade-offs.

\begin{figure*}[htb]
    \centering
    \includegraphics[width=\linewidth]{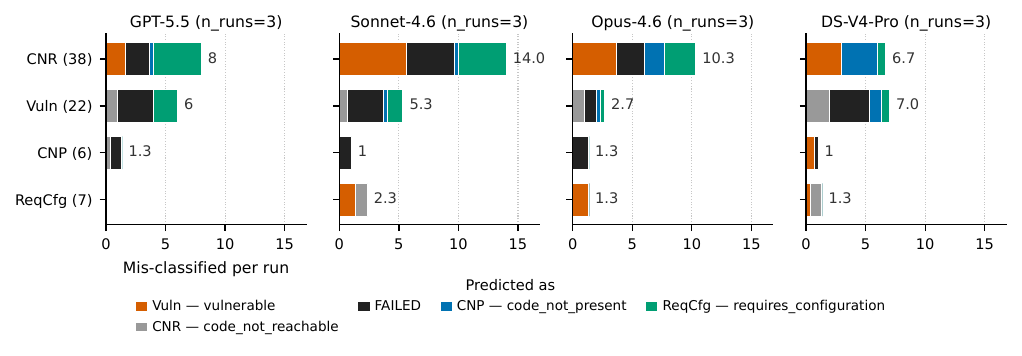}
    \caption{Error breakdown by ground-truth output category for four representative configurations. Each bar shows the average number of incorrect predictions per run, stacked by the category predicted by the agent. Categories with fewer than three benchmark cases are omitted.}
    \label{fig:errors-by-category}
\end{figure*}

\subsubsection{Language Analysis}

\begin{figure}[!h]
  \centering
  \includegraphics[width=0.9\linewidth]{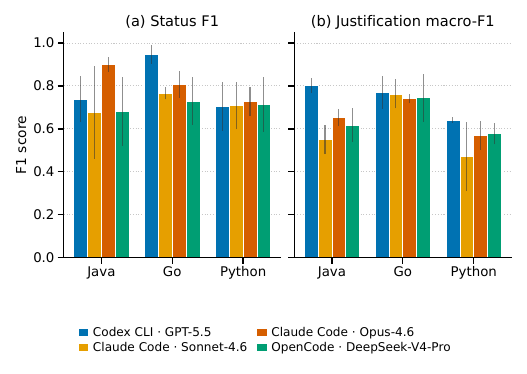}
  \caption{Per-language F1 for four representative configurations. (a) Status F1. (b) Justification macro-F1.}
  \label{fig:f1_by_lang}
\end{figure}

\textbf{Performance varies across programming languages.} \Cref{fig:f1_by_lang} breaks down performance by language for four representative configurations. Averaged over these configurations, status F1 ranges from 71.3\% on Python to 81.3\% on Go, a 10.0-point spread. In contrast, justification macro-F1 ranges from 56.5\% on Python to 75.5\% on Go, a 19.0-point spread. Go achieves the highest average justification macro-F1, while Python is the weakest language on both justification macro-F1 and average status F1. This suggests that language ecosystems affect agents' ability to identify the correct exploitability reason. One possible explanation is that Go has a comparatively mature and manageable package ecosystem, whereas Java projects often involve more complex multi-module dependency structures and Python projects frequently rely on loosely specified dependencies.

\subsubsection{Error Analysis}

\textbf{Models share where they fail but differ in how they fail.} \Cref{fig:errors-by-category} shows that errors across representative configurations concentrate in \texttt{code\_not\_reachable} (CNR) and \texttt{vulnerable} cases, yet the misclassification patterns diverge by model. \texttt{FAILED} denotes runs that do not produce a valid, evaluable output because of a timeout, a parsing failure, or a prediction outside the five annotated classes. GPT-5.5 frequently predicts \texttt{requires\_configuration} for CNR cases. We manually reviewed the five cases with this error pattern and found that four admit both explanations: the default configuration does not activate a path to the vulnerable functionality, whereas a non-default configuration can activate the corresponding component. Under our precedence rule, these cases are labeled \texttt{code\_not\_reachable}; nevertheless, \texttt{requires\_configuration} captures an additional condition supported by the evidence. These predictions therefore violate the required precedence but may still be factually valid. Claude Opus~4.6, in contrast, often over-calls CNR cases as \texttt{vulnerable}, indicating a more aggressive tendency to treat dependency evidence as exploitable risk. DeepSeek-V4-Pro exhibits a more balanced but less decisive profile, with comparable error counts on \texttt{vulnerable} and CNR cases, suggesting an unstable boundary between truly affected cases and non-exploitable dependency paths. Across models, errors cluster near the operational boundary between filtering false dependency alerts and identifying truly affected projects.

\subsubsection{Repository Size Analysis}

\begin{figure}[!t]
    \centering
    \includegraphics[width=\linewidth]{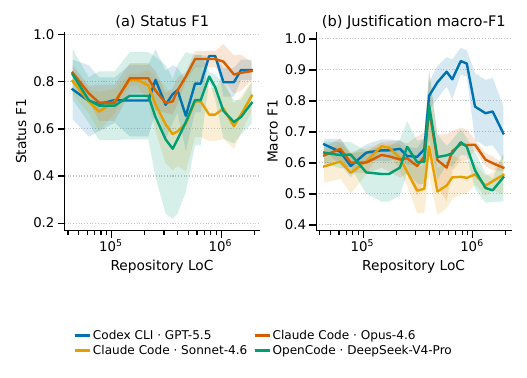}
    \caption{Status F1 and justification macro-F1 as a function of target repository size for four representative configurations. Each point is computed over a 25-task sliding window sorted by language-specific repository LOC. Shaded regions show run-to-run standard deviation when available.}
    \label{fig:f1-vs-repo-loc}
\end{figure}

\textbf{Repository size is not the primary bottleneck for strong agents, but it makes fine-grained reasoning less stable.} \Cref{fig:f1-vs-repo-loc} examines 25-task sliding windows ranging from about 44K to 1.9M lines of code. GPT-5.5 and Claude Opus~4.6 maintain high status F1 in the largest windows, reaching 85.0\% and 84.5\%, respectively. DeepSeek-V4-Pro is more sensitive to repository size, dropping from 83.2\% status F1 in the smallest window to 71.0\% in the largest window. Across models, the justification macro-F1 curves are lower and more variable than the status-F1 curves, especially in large repositories. This indicates that scaling to larger codebases mainly stresses the agent's ability to localize and explain the relevant dependency path, rather than only its ability to make a binary affected/not-affected decision.

\subsubsection{Harness Ablation Study}

\begin{figure}[!t]
    \centering
    \includegraphics[width=\linewidth]{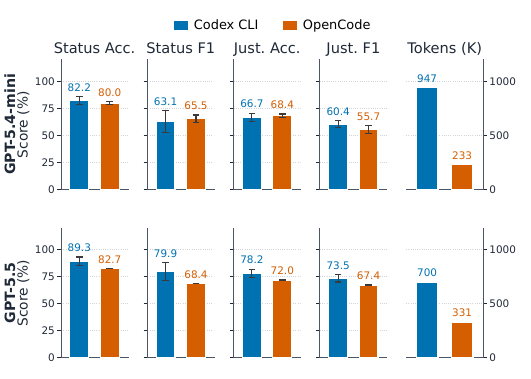}

    \caption{Ablation on the harness for GPT-5.4 mini and GPT-5.5 on \textsc{VEX-Bench}. Each row shows one model evaluated under the Codex CLI and OpenCode harnesses, reporting vulnerability-status accuracy/F1, justification accuracy/F1, and average per-case token usage (in thousands).}
\label{fig:harness-comparison}
\end{figure}

\textbf{Harness frameworks yield modest performance differences but substantially different token costs.} \Cref{fig:harness-comparison} compares GPT-5.4~mini and GPT-5.5 under Codex CLI and OpenCode. For GPT-5.5, Codex CLI consistently outperforms OpenCode across all metrics, but at the cost of more than doubling per-case token usage. For GPT-5.4~mini, the two harnesses are comparable in performance, yet Codex CLI consumes roughly 4$\times$ more tokens. Overall, harness choice affects inference cost far more than classification performance.


\section{Conclusion}
We introduced \textsc{VEX-Bench}, the first benchmark designed to evaluate whether LLM agents can assess the exploitability of software supply chain vulnerabilities. The benchmark comprises 75 real-world cases drawn from Go, Python, and Java projects, with ground-truth labels manually annotated by security experts. Our evaluation of nine model--harness configurations shows that these systems perform better on binary vulnerability-status classification than on the more fine-grained task of determining justification status. Overall, the results suggest that LLM agents are effective at identifying whether a project is affected by a vulnerability but still lack the depth of analysis required to provide accurate justifications.

\section*{Limitations}

\textsc{VEX-Bench} has three main limitations. First, its scale and coverage
are constrained by the high cost of expert annotation. The benchmark contains
75 cases across three language ecosystems, and some fine-grained justification
categories contain only a small number of examples. Moreover, the current
dataset covers only four types of not-affected justifications. This dataset also does not include other widely used ecosystems, such as JavaScript and npm. Consequently,
the reported results may not fully capture the diversity of exploitability
assessment challenges encountered in practice, particularly at the
fine-grained justification level.

Second, a single case may admit multiple valid explanations. We apply a predefined precedence rule to assign one gold label
and make evaluation unambiguous. Nevertheless, this single-label formulation
may omit valid secondary reasons for why a project is not affected.

Third, \textsc{VEX-Bench} assesses repository snapshots using the source code
and configuration information available in the repository. Such static
evidence cannot fully capture dynamic user inputs or deployment-specific
environment settings. The resulting judgments therefore approximate
exploitability under the repository's documented or default conditions and may
not reflect every real-world deployment.

\section*{Ethical Considerations}
Research on vulnerability exploitability has an inherent dual-use risk. In
principle, the ability to determine whether a vulnerable dependency is
reachable in a downstream project could help malicious actors prioritize known
vulnerabilities for further investigation. However, \textsc{VEX-Bench} uses
only publicly disclosed CVEs and does not involve zero-day discovery. Its tasks
ask agents to classify exploitability and justify their decisions with code and
advisory evidence; they do not require exploit generation, payload construction,
or attacks against running services. In addition, all experiments analyze
source-code snapshots inside isolated Docker environments without access to
deployed systems, reducing their direct operational value for attacks and the
risk of affecting real-world infrastructure. \textsc{VEX-Bench} is intended for defensive vulnerability triage, helping defenders prioritize genuine risks, reduce false-positive alerts, and evaluate the capabilities of LLM agents.

\section*{Acknowledgments}
We sincerely thank the anonymous reviewers for their constructive feedback. We also thank Theodor Mihalache, Shimon Tanny, Tamar Weisskopf, Michelle DiPalma, and Jude Niroshan at Red Hat for their contributions to this work. This work was supported in part by NSF Proto-OKN Award 2333736.

\bibliography{refs}

\appendix

\section{Justification Category Definitions}
\label{appendix:justification}

\textsc{VEX-Bench} reports four justification categories that explain why a
downstream project is \emph{not} affected by a known vulnerability in one of
its dependencies. \Cref{tab:justification-categories} summarizes the four
categories at a glance; the paragraphs that follow describe, for each
category, the underlying condition, the evidence required to assign it, and
a representative case from \textsc{VEX-Bench}.

\begin{table*}[h]
\centering
\small
\caption{The four \texttt{not\_affected} justification categories used in
\textsc{VEX-Bench}, adapted from the VEX status-justification
vocabulary~\citep{cisa2022vexstatus}.}
\label{tab:justification-categories}
\begin{tabular}{p{0.22\linewidth} p{0.74\linewidth}}
\toprule
\textbf{Category} & \textbf{Definition} \\
\midrule
\texttt{code\_not\_present}
  & The vulnerable package or module is absent from the project's resolved
    dependency graph: not declared in any manifest, not pinned in any
    lockfile, not vendored, and not pulled in transitively. \\
\addlinespace
\texttt{code\_not\_reachable}
  & The vulnerable function exists somewhere in the dependency tree but is
    never invoked along any reachable call path from first-party code. \\
\addlinespace
\texttt{requires\_configuration}
  & The vulnerable code path is reachable, but exploitation requires a
    non-default configuration option, feature flag, or runtime setting that
    the project does not enable. \\
\addlinespace
\texttt{requires\_environment}
  & The vulnerable code is reachable, but its trigger depends on a runtime
    environment (operating system, kernel feature, hardware capability, or
    external service) that the project does not establish. \\
\bottomrule
\end{tabular}
\end{table*}

Although \textsc{VEX-Bench} reports four justification categories in the main
benchmark, we considered a broader VEX-aligned category space when designing
the task prompt. The additional categories listed in
\Cref{tab:justification-categories-future} represent theoretically possible
not-affected reasons and boundary cases drawn from the status-justification
vocabulary maintained by CISA for VEX~\citep{cisa2022vexstatus}. They document
the prompt design, but are not introduced as separate benchmark classes. All
dataset statistics and justification metrics in the paper are reported over
the four categories above.

\begin{table*}[h]
\centering
\small
\caption{Broader VEX-aligned categories considered during prompt construction.
These categories represent theoretically possible not-affected reasons and
boundary cases, but are not separate benchmark classes in the reported
evaluation.}
\label{tab:justification-categories-future}
\begin{tabular}{p{0.26\linewidth} p{0.70\linewidth}}
\toprule
\textbf{Category} & \textbf{Definition} \\
\midrule
\texttt{requires\_dependency}
  & Exploitation requires an additional optional dependency (library,
    plugin, or module) that the project does not declare. \\
\addlinespace
\texttt{compiler\_protected}
  & Compile-time hardening configured in the project's build pipeline
    (stack canaries, PIE/ASLR-enforced builds, sanitizer instrumentation)
    prevents the exploit primitive. \\
\addlinespace
\texttt{runtime\_protected}
  & Runtime mechanisms instantiated by the project's own code (sandboxing,
    in-process isolation, seccomp filters) prevent exploitation. \\
\addlinespace
\texttt{perimeter\_protected}
  & Network, authentication, or perimeter controls shipped with the
    project (authentication middleware, allowlists, deployment-level
    network policies) block the attack surface. \\
\addlinespace
\texttt{mitigating\_control\_protected}
  & Other in-repository safeguards not covered above (input validation,
    output encoding, custom guards) reduce risk to negligible. \\
\bottomrule
\end{tabular}
\end{table*}

\paragraph{\texttt{code\_not\_present}}
The vulnerable package or module is absent from the project's resolved
dependency graph: not declared in any manifest, not pinned in any lockfile,
not vendored, and not pulled in transitively. This label is assigned only
after inspecting every dependency manifest, the corresponding lockfiles, and
any vendored or generated sources, since a vulnerability database entry for
the broader package family does not by itself imply that the affected module
is consumed by this project. \emph{Example.} CVE-2024-7254 in
\texttt{apache/dubbo} at commit \texttt{218aab1}: although the underlying
library appears in upstream advisories, the affected module is not part of
the project's resolved dependencies, so no call path can reach the
vulnerable code.

\paragraph{\texttt{code\_not\_reachable}}
The vulnerable function is present somewhere in the dependency tree but is
never invoked along any reachable call path from first-party code. The
vulnerable surface may live in a transitive dependency, in an unused
submodule of a direct dependency, or behind an import path that the project
does not exercise. Assigning this label requires identifying the vulnerable
functions named in the advisory or its fix and verifying that no import,
reflection edge, or call site in first-party source leads to them.
\emph{Example.} CVE-2025-55752 in \texttt{alibaba/nacos} at commit
\texttt{20caf99}: the affected library is included in the dependency graph,
but the specific APIs that the patch modifies are not referenced anywhere in
the project's source.

\paragraph{\texttt{requires\_configuration}}
The vulnerable code path exists and is reachable, but exploitation requires
a non-default configuration option, feature flag, or runtime setting that
the project does not enable. The controlling configuration may live in a
YAML or properties file shipped with the project, an environment variable
read at startup, or a hard-coded constant in first-party code. Assigning
this label requires locating the controlling configuration source,
confirming that its disabling value is in effect by default, and checking
that no other code path overrides it at runtime. \emph{Example.}
CVE-2023-47248 in \texttt{apache/airflow} at commit \texttt{3dbad4f}: the
vulnerable serialization behavior is gated by a configuration switch whose
default value disables the affected functionality.

\paragraph{\texttt{requires\_environment}}
The vulnerable code is reachable but its trigger depends on a runtime
environment that the project does not establish. The required environment
may be an operating system, kernel feature, hardware capability, or
external service that the project's deployment manifests, Dockerfiles, or
installation instructions do not provide. Assigning this label requires
reading the advisory's precondition description and confirming that the
host project's declared environment lies outside the affected range.
\emph{Example.} CVE-2024-47535 in \texttt{prestodb/presto} at commit
\texttt{8136f7c}: the exploit's required runtime environment is not part
of the project's supported deployment surface.

\section{Prompts}
\label{sec:prompts}

\subsection{Task Instruction Prompt}
\label{app:task-inst-prompt}
The prompt below reflects the broader VEX-aligned category space considered
during prompt construction. It should be read as the instruction template used
to make the agent's decision process explicit; the benchmark label space
reported in the paper remains the four categories defined in
\Cref{tab:justification-categories}.

\begin{lstlisting}[style=prompt, caption={This is task instruction prompt template.}, label=lst:task-inst-prompt]
<ROLE>
You are a security analyst. Decide whether the given CVE actually affects current codebase, and classify the result into one of 12 fixed categories.
</ROLE>

<CONTEXT>
The working directory is a source tree. There is no running container, no running process, and no deployment context: only files you can read and search, plus any public CVE/advisory information you look up.

Treat this as a static analysis task. Do not assume runtime behavior you cannot back with evidence visible in the source tree.
</CONTEXT>

<CVE>
{cve_id}
</CVE>

<CLASSIFICATION_CATEGORIES>
The 12 categories are listed below in STRICT LOGICAL PRECEDENCE ORDER. When evaluating, walk through the list from top to bottom and select the FIRST category that applies. Do not skip ahead.

1. "false_positive"
   - The CVE-to-package mapping is wrong (named package is not what the CVE applies to, or the CVE is withdrawn/malformed). Use only with concrete evidence of mismatch.
   
2. "code_not_present"
   - The vulnerable package/module is absent from the repository: not declared in any manifest, not in any lockfile, no vendored copy.

3. "code_not_reachable"
   - The vulnerable code is present (manifest, lockfile, or vendored copy) in the codebase but is never executed at runtime, e.g., never imported, referenced, or called from first-party source. 
   - Only applicable when code IS present AND call-chain/reachability analysis confirms no execution path leads to it.

4. "requires_configuration"
   - Exploitation requires a specific configuration option, feature flag, or setting that is currently disabled by default in this repository.

5. "requires_dependency"
   - Exploitation requires an additional dependency (library, plugin, module) that this repository does not declare.

6. "requires_environment"
   - Exploitation requires a specific runtime environment (OS, architecture, kernel version, hardware feature) that this repository does not establish or rely on.

7. "compiler_protected"
   - Compile-time hardening configured in this repository's build (stack canaries, PIE/ASLR-required builds, sanitizer instrumentation, etc.) prevents the exploit primitive.

8. "runtime_protected"
   - Runtime mechanisms set up by this repository's own code (sandboxing, in-process isolation, seccomp filters) prevent exploitation.

9. "perimeter_protected"
   - Network, authentication, or perimeter controls shipped in this repository (auth middleware, allowlists, network policies in deployment manifests) block the attack surface.

10. "mitigating_control_protected"
    - Other in-repo mitigations not covered by 7-9 (input validation, output encoding, custom guards) reduce risk to negligible.

11. "uncertain"
    - Investigation cannot establish presence, reachability, or mitigation status with the available evidence. Use as a true fallback, not a hedge.

12. "vulnerable" 
   - All of:
    - an affected version of the vulnerable package is present,
    - the vulnerable surface is imported and called from first-party (non-test) source,
    - no mitigation from categories 4-10 applies.
</CLASSIFICATION_CATEGORIES>

<DECISION_RULES>
1. A CVE is classified as "vulnerable" if and only if ALL of the following hold:
   - The vulnerable code is PRESENT in the container/codebase.
   - The vulnerable code is USED or CALLED by the application.
   - The vulnerable code is REACHABLE from an attack surface (user input, network input, file processing, IPC, etc.).
   - No effective mitigations or protections are in place.

2. If ANY of the above conditions fails, select the SINGLE most appropriate non-vulnerable category by walking the precedence list from top (1) to bottom (11) and choosing the first matching category. For example:
   - If the vulnerable code is not present → "code_not_present" (do NOT also consider "code_not_reachable" or environment factors).
   - If the code is present but unreachable → "code_not_reachable" (do NOT fall through to "requires_environment").
   - If a required dependency is missing → "requires_dependency".
   - If the vulnerable code is prevented by a default or clearly set configuration → "requires_configuration".

3. Use "uncertain" only when the investigation genuinely lacks the evidence needed to reach any conclusion.
</DECISION_RULES>

<OUTPUT_FORMAT>
Output a single JSON object on stdout — no surrounding prose, no markdown fences, no comments.

Schema:
{
  "category": "<one of the 12 category names, exact snake_case>",
  "reasoning": "<evidence-based explanation; multi-line strings are fine>"
}

Reasoning must be grounded in concrete evidence — cite file paths, manifest entries, version numbers, function names, or advisory fragments. Be specific; avoid vague claims like "the code looks safe". Newlines inside the reasoning string must be escaped (`\\n`) to keep the object valid JSON.
</OUTPUT_FORMAT>

<EXAMPLE>
{"category": "code_not_reachable", "reasoning": "GHSA-44wm-f244-xhp3 describes a vulnerability in PIL.ImageMath.eval for Pillow < 10.3.0. pyproject.toml pins Pillow to ^9.5, and poetry.lock records 9.5.0 as the resolved version — within the affected range, so the vulnerable code is in scope.\\n\\nSearching first-party source (`rg \\\"ImageMath\\\" src/`) returns no matches. The repository imports PIL.Image only, and the only call sites are `Image.open()` and `Image.thumbnail()` in src/img/loader.py:14-37. Neither reaches ImageMath.eval, so the vulnerable function is unreachable."}
</EXAMPLE>
\end{lstlisting}

\paragraph{Mapping to Benchmark Labels.}
Although the prompt allows 12 categories, \textsc{VEX-Bench} evaluates only
five annotated classes: \texttt{vulnerable} and the four not-affected
categories defined in Appendix~\ref{appendix:justification}. A
\texttt{vulnerable} prediction maps to the affected status, while the four
annotated justification categories map to the not-affected status. Predictions
in any of the remaining categories are marked as \texttt{FAILED}; they are
neither remapped to an annotated class nor excluded from evaluation.

\section{Model-Side Leakage Analysis}
\label{app:model-side-leakage}

Because \textsc{VEX-Bench} is constructed from public repositories, the source
PRs may have appeared in model training data even though the benchmark labels
are not public. We assess this potential source of leakage through a manual
audit of the source PRs and a guided-completion probe for memorized PR content.

\paragraph{Source PR Audit}
We manually review the titles, descriptions, and discussion threads of the
source PRs associated with all 75 cases. Only three discuss whether the
corresponding vulnerability is exploitable, and none contains the vulnerability
status or justification assigned by our annotators. The public source PRs
therefore do not directly reveal the benchmark answers.

\paragraph{Guided-Completion Probe}
Following prior work~\citep{golchin2024time}, we further probe whether the
evaluated models reproduce source PR content from memory. We select PRs that
were created before the models' knowledge cutoff
dates, written by humans rather than bots, non-boilerplate, and sufficiently
long to support a meaningful completion test. Nineteen distinct PRs satisfy
these criteria. We test all of them with GPT-5.5 and Claude Opus~4.6, with web
access and tools disabled, and classify each response as an exact match, a
verified near-exact match, a nonmatching completion, or a refusal/empty
response.

\begin{table*}[h]
\centering
\small
\caption{Results of the guided-completion probe on the 19 source PRs selected
for model-side leakage analysis.}
\label{tab:model-side-leakage}
\begin{tabular}{lrrrr}
\toprule
\textbf{Model} & \textbf{Exact} & \textbf{Verified Near-exact} &
\textbf{Nonmatching} & \textbf{Refusal/Empty} \\
\midrule
GPT-5.5 ($n=19$)            & 1 & 0 & 12 & 6 \\
Claude Opus~4.6 ($n=19$)   & 0 & 0 & 19 & 0 \\
\midrule
Combined ($n=38$)          & 1 & 0 & 31 & 6 \\
\bottomrule
\end{tabular}
\end{table*}

Across the 38 completions, we observe no verified near-exact matches and only
one exact ten-token continuation. This match occurs because the PR body repeats
its title, which is already provided in the prompt, rather than because the
model reproduces unseen PR content. Together with the source PR audit, these
results provide no meaningful evidence that the evaluated models memorized the
source PR contents or could recover the benchmark answers from them.

\section{Experimental Details}
\label{app:experimental-details}

\subsection{Environment}
\begin{table*}[h]
  \centering
  \small
  \caption{Language runtime environments used in our Docker containers.}
  \label{tab:language-env}
  \begin{tabular}{llll}
  \toprule
  Language & Base image & OS & Runtime / Build tool \\
  \midrule
  Python & \texttt{python:3.13-slim-trixie} & Debian GNU/Linux 13 & Python \texttt{3.13.13} \\
  Go & \texttt{golang:1.26-bookworm} & Debian GNU/Linux 12 & Go \texttt{1.26.3} \\
  Java & \texttt{eclipse-temurin:21-jdk} & Ubuntu 26.04 LTS & OpenJDK \texttt{21.0.11}; Maven \texttt{3.9.12} \\
  \bottomrule
  \end{tabular}
\end{table*}
Table~\ref{tab:language-env} summarizes the language-level runtime environments used in our Docker containers. We report only the base
image, operating system, and primary runtime or build tool versions; common utilities such as \texttt{git}, \texttt{curl}, and
\texttt{ca-certificates} are inherited or installed as part of the container setup.

\subsection{Model Backends}
\label{app:model-backends}

All model API calls in our experiments were routed through Azure's model-serving
APIs. We used the Azure AI Model Inference REST API as the common provider
interface for deployed foundation models~\citep{microsoft2026azuremodelinference}.
Table~\ref{tab:model-backends} lists the model names used in the paper, the
corresponding provider-facing model card or release source, and the public
release date of each model.

\begin{table*}[ht]
  \centering
  \small
  \caption{Model backends used in our experiments.}
  \label{tab:model-backends}
  \begin{tabular}{p{0.24\linewidth}p{0.46\linewidth}p{0.20\linewidth}}
  \toprule
  Model name & Model card / source name & Release date \\
  \midrule
  Claude Opus~4.6 & Claude Opus 4.6~\citep{anthropic2026opus46} & Feb. 5, 2026 \\
  Claude Sonnet~4.6 & Claude Sonnet 4.6~\citep{anthropic2026sonnet46} & Feb. 17, 2026 \\
  GPT-5.5 & Introducing GPT-5.5~\citep{openai2026gpt55} & Apr. 23, 2026 \\
  GPT-5.4~mini & GPT-5.4 mini~\citep{openai2026gpt54mini} & Mar. 17, 2026 \\
  Kimi~K2.6 & Kimi K2.6~\citep{moonshot2026kimik26} & Apr. 20, 2026 \\
  DeepSeek-V4-Pro & DeepSeek-V4-Pro~\citep{deepseek2026v4} & Apr. 24, 2026 \\
  DeepSeek-V4-Flash & DeepSeek-V4-Flash~\citep{deepseek2026v4} & Apr. 24, 2026 \\
  MiniMax-M2.7 & MiniMax M2.7~\citep{minimax2026m27} & Mar. 18, 2026 \\
  GLM-5.1 & GLM-5.1~\citep{zai2026glm51} & Apr. 7, 2026 \\
  \bottomrule
  \end{tabular}
\end{table*}

\subsection{Harness Framework}
\label{app:harness-framework}

\begin{table*}[h]
  \centering
  \small
  \caption{Agent harness versions and effort settings used in our experiments.}
  \label{tab:agent-env}
  \begin{tabular}{p{0.16\linewidth}p{0.42\linewidth}p{0.14\linewidth}p{0.18\linewidth}}
  \toprule
  Agent & Website & Version & Effort setting \\
  \midrule
  Codex & \url{https://developers.openai.com/codex/cli} & \texttt{0.131.0} & \texttt{xhigh} \\
  Claude Code & \url{https://code.claude.com/docs/en/setup} & \texttt{2.1.150} & \texttt{high} \\
  OpenCode & \url{https://opencode.ai/docs/} & \texttt{1.15.5} & Not set \\
  \bottomrule
  \end{tabular}
\end{table*}

Table~\ref{tab:agent-env} reports the harness frameworks used in our
experiments. For each agent, we list the corresponding official website or
documentation page, the exact CLI version installed in the experimental
containers, and the effort setting used when the harness exposes such a
control. We used the strongest available effort setting in our setup: Codex was
run with \texttt{xhigh}, and Claude Code was run with \texttt{high}. For the
Claude~4.6 models, \texttt{high} is the maximum supported effort level. OpenCode
was run without an effort override because no analogous effort setting was
configured in our experiments. These versions and settings identify the
agent-side execution environment; model identifiers and provider backends are
reported separately from the tool versions.

\subsection{Runtime and Timeout Analysis}
\label{app:runtime-analysis}

\begin{table*}[!ht]
  \centering
  \small
  \caption{End-to-end wall-clock runtime across all executions. Median, P90, and maximum runtimes are reported in seconds. Timeout is the percentage of executions that reach the 600-second limit; consequently, a maximum of 600 seconds indicates at least one timed-out execution.}
  \label{tab:runtime-statistics}
  \begin{tabular}{lrrrr}
  \toprule
  Model & Median (s) & P90 (s) & Max (s) & Timeout (\%) \\
  \midrule
  Claude Opus~4.6 & 105 & 217 & 336 & 0.0 \\
  Claude Sonnet~4.6 & 108 & 177 & 456 & 0.0 \\
  GPT-5.5 & 162 & 253 & 442 & 0.0 \\
  GPT-5.4~mini & 176 & 256 & 380 & 0.0 \\
  GLM-5.1 & 79 & 246 & 562 & 0.0 \\
  DeepSeek-V4-Pro & 99 & 215 & 600 & 0.4 \\
  Kimi~K2.6 & 135 & 364 & 600 & 3.1 \\
  DeepSeek-V4-Flash & 151 & 456 & 600 & 0.4 \\
  MiniMax-M2.7 & 91 & 600 & 600 & 12.0 \\
  \bottomrule
  \end{tabular}
\end{table*}

Overall, 98.2\% of executions complete within the 10-minute limit, indicating
that the cap accommodates most agent trajectories. Runtime nevertheless varies
substantially across configurations, particularly in the upper tail. Most
timeouts occur for Kimi~K2.6 and MiniMax-M2.7. Manual inspection shows that
Kimi's timeouts mainly arise from long analysis trajectories on large
repositories, whereas MiniMax's timeouts are primarily associated with slow API
responses. Because these measurements cover the complete execution, they
reflect the combined latency of the model backend, agent harness, and tool-use
trajectory rather than model inference alone.

\section{Responsible Research Details}
\label{app:responsible-research}

\subsection{Artifact Use and Intended Use}
\label{app:artifact-use}

\textsc{VEX-Bench} is constructed from public open-source repositories and
public vulnerability records for research evaluation of software supply chain
vulnerability triage. This use is consistent with the public and research-facing
nature of the source artifacts: the repositories are selected from permissively
licensed open-source projects, and the vulnerability identifiers and advisories
describe already disclosed security issues. The benchmark artifacts we create,
including task metadata, labels, prompts, and evaluation scripts, are intended
for defensive triage research, evaluation of LLM-agent reliability, and analysis
of failure modes. They are not intended to support exploit generation,
unauthorized vulnerability testing, or attacks against deployed systems. Users
of the benchmark are responsible for complying with the licenses and terms of
the upstream repositories, tools, models, and vulnerability-data sources they
use.

\subsection{Data Privacy and Offensive Content}
\label{app:data-privacy}

The benchmark is derived from public source-code repositories and public
vulnerability records rather than from private user data. During curation, we do
not collect user profiles, private communications, issue discussions, code-review
threads, or other person-centered content. The code snapshots used in evaluation
remove Git history, which avoids retaining commit-author timelines and email
metadata. The released benchmark metadata is intended to contain task
identifiers, repository references, CVE identifiers, labels, annotations,
prompts, and evaluation scripts, rather than personal information about project
contributors or annotators.

We do not intentionally collect offensive content. Because the source artifacts
are real software projects and public security advisories, they may contain
security-related terminology or public project/license notices, but such content
is included only insofar as it is necessary for vulnerability triage research.
Before release, we review benchmark metadata to avoid including unnecessary
personal identifiers or unrelated sensitive content.

\subsection{Potential Risks}
\label{app:potential-risks}

This work studies software supply chain vulnerability exploitability, so it has
an inherent dual-use aspect. A benchmark that evaluates whether agents can reason
about vulnerable dependencies and downstream reachability could, in principle,
help a malicious actor prioritize which known vulnerabilities deserve further
manual investigation. However, our work is designed as a defensive evaluation
resource rather than an exploitation system. The tasks are based on already
disclosed vulnerabilities, and the benchmark asks agents to determine
exploitability status and provide evidence for the decision; it does not require
agents to generate exploit payloads, compromise running services, or perform
end-to-end attacks.

We further reduce operational risk by framing the task as static analysis of
source-code snapshots. The experimental environments are isolated containers
without deployment context, and the prompts explicitly instruct agents to ground
their answers in code and advisory evidence rather than to execute attacks. The
intended use of \textsc{VEX-Bench} is to improve defensive triage of dependency
alerts, help researchers measure the reliability of LLM agents for security
analysis, and identify current failure modes before such agents are used in
high-stakes software maintenance workflows.

\subsection{Annotation Instructions and Consent}
\label{app:annotation-instructions}

Annotators were instructed to determine whether a downstream project is actually
affected by a known vulnerability in one of its third-party dependencies. For
each candidate case, annotators were given the target repository snapshot, the
relevant CVE identifier, and the dependency-update pull request from which the
case was mined. They were asked to inspect the dependency manifest or lockfile,
the public vulnerability advisory, the upstream patch when available, and the
downstream project code needed to assess whether the vulnerable functionality is
present, called, and reachable from the project.

The annotation task had two outputs. First, annotators assigned a binary
vulnerability status indicating whether the downstream project is affected by
the dependency vulnerability. Second, annotators assigned a justification label
explaining the status. A case should be labeled as affected only when the
vulnerable dependency version is present, the vulnerable functionality is used by
the downstream project, the relevant code path is reachable from a plausible
input or execution path, and no in-repository mitigation blocks exploitation. If
any of these conditions is not met, annotators selected the most specific
not-affected justification in the reported benchmark label space. These four
categories cover missing vulnerable code, unreachable code, required
configuration, and required environment. The additional categories shown in the
prompt are part of the broader design space used to formulate the task, not
additional reported benchmark classes.

Annotators were instructed to base labels on concrete evidence, including file
paths, dependency versions, function or class names, configuration values, and
advisory or patch details. They were asked not to infer exploitability from the
existence of a vulnerable dependency alone, and not to label a case as affected
unless they could identify a downstream path that uses the vulnerable
functionality under the required conditions. Ambiguous cases were discussed
during calibration meetings, and final labels were assigned after reconciling
disagreements among annotators. No screenshot-based annotation interface was
used.

Annotators were members of the research team and were informed that their
annotations would be used to construct the benchmark labels and to report
aggregate dataset statistics. We do not release annotator identities,
annotator-level performance, or other personal information about annotators.

\subsection{Use of AI Assistants}
\label{app:ai-assistants}

The authors used AI assistants for language polishing, manuscript revision, and
literature-search assistance. All AI-generated suggestions were reviewed,
edited, and verified by the authors. AI assistants were not used to create
ground-truth labels, conduct final exploitability judgments, generate
experimental results, or make unverified scientific claims.

\end{document}